# THE COOPERATIVE JAHN-TELLER ORDERING INVESTIGATION BY ABSORPTION SPECTRA IN $KDy(MoO_4)_2$

N.M. Nesterenko, Yu. Kharchenko

B. Verkin Institute for Low Temperature Physics and Engineering of the National Academy of Sciences of Ukraine

Absorption spectra fine structure of $KDy(MoO_4)_2$ in the region of cooperative Jahn - Teller type ordering was studied. Temperature anomalies in the spectra occurring at phase transformation correlate with the ultrasound peculiarities observed earlier. Based on the symmetry approaching, possible activity of the irreducible representations of the rhombic $D_{2h}$ point group was discussed, which lead to the incommensurate phase at cooperative ordering. It was supposed, that coupled $A_u$ – type phonon mode may lead to the incommensurate phase existence, which is possible at least in the temperature region 17-12K.

## I. INTRODUCTION.

The second - order phase transition in $KDy(MoO_4)_2$ was found and described firstly in [1,2]. It was attributed to the pseudo-cooperative Jahn - Teller effect (CJTE) at ~ 15K. When the birefringence and magnetic properties studies of $KDy(MoO_4)_2$ were performed [3,4], it was shown that a sequence of the phase transitions occurs during CJTE with transition temperatures $T_1^{cr}$ = 14.5K and $T_2^{cr}$ = 11.5K. The phase, which exists in the crystal between $T_1^{cr}$ and $T_2^{cr}$, was assumed in [3,4] as the incommensurate one. It was shown earlier by symmetry approach in [5, 6, 7, 8], that ultrasound anomalies are compatible with the activity of Brillouin zone boundary point $\mathbf{k}_{23}$ = $1/2(\mathbf{b}_2+\mathbf{b}_3)$ at the structure phase transition (notations for $D_{2h}^{14}$ space group are the same as in [9]), and only the monoclinic commensurate phase is induced in $KDy(MoO_4)_2$ due to the second order phase transition. More detailed examination of the experimental data show, however, that ultrasound anomalies are different in the temperature regions near $T_1$ ~ 12K and $T_2$ ~ 17K [5, 6, 8, 10]. Note, that to argue the zone Brillouin point activity, the authors [5] took into account ultrasound anomalies just at 12K. This fact does not contradict to the supposition that the intermediate phase may exist in the temperature region 17-12K. So the question about the symmetry and physical origin of the phase which exists between $T_1$ and $T_2$ (or $T_1^{cr}$ and $T_2^{cr}$) continues to be actual.

The aim of this work was to study the temperature dependencies of the fine structure absorption spectra of KDy – molybdate, and to detect its peculiar features in the temperature region 30-2K. Other task was to determine the correlations in the temperature behavior of the elastic modules and the absorption spectra parameters. Basing on symmetry analyses it was shown the possible way of the CJTE realisation in this crystal.

The absorption spectra were recorded at the energies 13200 – 1300 cm$^{-1}$ and 12200 - 12250 cm$^{-1}$ with a device based on a DFS-12 double monochromator. Samples of KDy(MoO$_4$)$_2$ were plates of 0.1 - 1 mm thick, the natural shear plane is the {010} surface.

## II. EXPERIMENT.

The absorption spectra fine structure of KDy(MoO$_4$)$_2$ in the studied energy intervals are caused by transitions from the components of the ground state $^6H_{15/2}$, which is split by the rhombic crystal field to 8 components, to the excited states $^6F_{5/2}$ and $^6F_{3/2}$ within f-configuration of Dy$^{3+}$ ions. Since the samples were the plates, the spectra were recorded for polarisation **E**||**a**, **E**||**c** only, the parameters **a** and **c** of the rhombic phase are within the shear plane.

The temperature evolution of the spectrum induced by the $^6H_{15/2}$ - $^6F_{3/2}$ transition, is shown for **E**||**a** in Figure 1. At 2K and 6K the doublets were observed for two polarisations of the light - **E**||**a** (Figure 1) and **E**||**c** - with the similar parameters of the bands. To clear up the doublet nature of the lines at 2K (6K), the absorption spectra were studied in the energy region 12200-12250cm$^{-1}$ ($^6H_{15/2}$-$^6F_{5/2}$ transition) at the same spectral gaps ~2cm$^{-1}$. At 2K spectrum was presented by 3 single bands, which had the different intensities, and every line was a single one, but not a doublet. The number of lines in the spectrum was in accordance with the maximum number of components for excited states $^6F_{5/2}$ due to it's splitting by rhombic field in the crystal (2J+1)/2 = 3, J=5/2. We therefore can conclude, that the doublet structure of the spectrum in Figure 1 is caused by the splitting of the excited state $^6F_{3/2}$, namely, on 7.5 – 8 cm$^{-1}$.

As the temperature rises, a new line appears at the low-energy side of the doublet and the energy distance between satellite and the initial doublet is reduced. If we take into account, that the first excited level is thermally occupied at the 14-19K, and the transition from excited level to $^6F_{3/2}$ may be presented, we can suppose that the doublet structure for satellite must be the same as for initial doublet (see scheme on Figure1). But it was not revealed generally. Temperature dependence of the energy interval between maximum of (0-0') component and satellite line maximum is shown on the Figure 2 (the notations are given in accordance with the scheme in Figure 1).

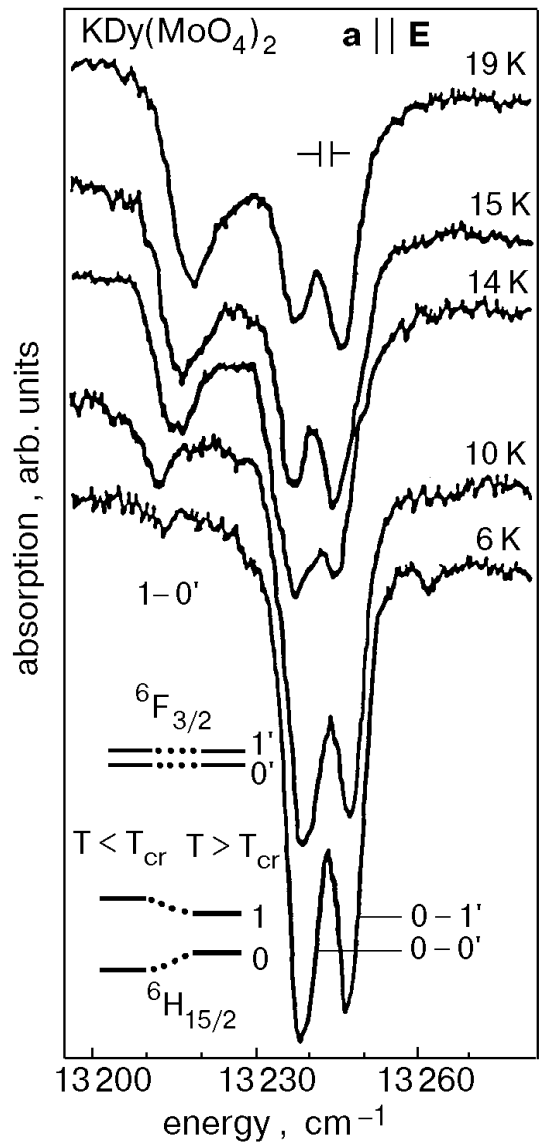

Figure 1. Temperature behaviour of the absorption spectra of KDy- molybdates at E∥a polarisation of the light, transition $^6H_{15/2}$- $^6F_{3/2}$ ; *a* is in the {010} plane. Scheme corresponds to the IR absorption spectra data for the one-ion representation.

We determined also, that when the temperature increased, the ratio of intensities of the components of the initial (low-temperature) doublet changed for studied polarisation. We have found, that the ratio of relative intensities $I_2/I_1$ underwent non-monotonic changing in the temperature region 2-20K, Figure 3, for every polarisation. Here $I_1$ is the peak intensity of the more intensive component (0-0') of the doublet at ~19K, Figure 1, the same for the both cases **E∥c** and

**E∥a.** So at T<12K the depolarisation of the doublet was observed, in contrast to the temperature region above 12K, when components of the doublet had different intensities in the both polarisations and the spectra were polarised.

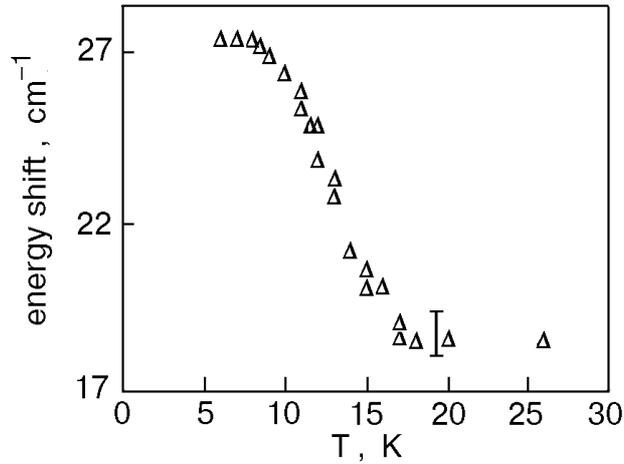

Figure 2. Temperature behaviour of the energy interval between 0-0' component and the satellite line for **E∥a**.

To exclude the accidental variations we have measured different samples and the result was the same.

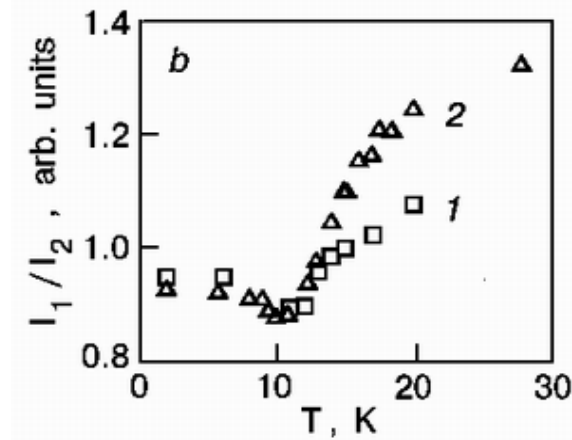

Figure 3. Ratio of intensities $I_1/I_2$ for **E∥c** (1) and **E∥a** (2), here $I_1$ and $I_2$ are the relative intensities of the doublet components, transition $^6F_{15/2} - ^6F_{3/2}$. For components 1 and 2 the ratio $I_1/I_2$ is less then 1 for low temperatures.

However, if we compare the scheme in the Figure 1 and the spectra at the finite temperature on the same Figure, we can see the contradiction, mentioned above: the absence of the doublet structure for arising line from the low-energy side. To understand this circumstance and the nature of the phase transition near 12K, we have applied the computer modelling of the spectra.

We have attempted computing the absorption spectrum at the low temperatures using the Lorentz – type form lines. The high – energy component of the spectra is not distorted due the other nearby lines, in contrast with the second doublet component, so we have used the line profile, corresponding to (0-1') transition. The half-width of the high-energy (0-1') component of the doublet was monotonously (linearly) changed from 6cm$^{-1}$ to 10cm$^{-1}$ when the temperature was changed from 2 to 25K. The same (fixed) values of half - width for the second component (0-0') of the doublet were used for the temperature higher than 12K. We do not observe noticeable widening of the lines in the region ~12K.

We have used the simplest models for computing. The Lorentz-type form of the lines is accepted for every component of the fitted spectra. The distance between doublets is changed according to Figure 2, and the intensities between doublets obey the Bose-factor.

We can see, that the changes in the spectra, caused by the thermal occupation of the exciting levels only, does not permit us to describe the evolution of the spectra. In other words, the intensity behaviour is more complicated and needs to take into account additional mechanism, which is imposed, partly, by phase transition at 12K. Besides, we determined, that the half-width of the high-energy component of the excited doublet is somewhat different from that we have observed in the experiment. The last spectrum (T=28K) was computed with the assumption that the half-width of the 1-0', 1-1' transitions is different from the same value for 0-0', 0-1' transitions, namely, they are 18 cm$^{-1}$. And, at last, we cannot describe the single line on the low-temperature side of the spectra only by this mechanism.

So, the main results of our experiment are the following. The energy shift between fine components of the lines (or between the energy levels) appears about $T_1$~17 K. We can describe its temperature dependence as the polynomial law. Non-monotonic "v"- type small changes of the line intensity were observed and at $T_2$ ~ 12K, the kink in it's temperature dependence takes place, and spectra became depolarised at T<$T_2$. By that way we have shown that the changes in the level position were observed at T>12K - above the phase transition temperature $T_2$ as it follows from [5, 8, 10]. The computing shows us, that the low-energy component of the spectra in the ordinary case, when the temperature is changed only, looks like the low-temperature good resolved doublet with the same interval 8 cm$^{-1}$ between two components.

We observed also a very weak line at 2K-6K at the distance about 18cm$^{-1}$ from the high – energy side of the (0-0') line, and this line does not correspond of the scheme (Figures 1 and 4). We cannot study its temperature dependence in detail, but we have shown, that this line was observed only at the very low temperatures, i.e. in the low-temperature phase.

## III. DISCUSSION.

### 3.1 ABSORPTION SPECTRA FORMATION.

We suppose, following [1], that the arising of the energy interval from 19 to 27 cm$^{-1}$ (Figure 4), when the temperature is lowering from 17 to 8 K, is the result of the lowering of the ground state of Jahn-Teller ions due to CJTE. Energy position of the first excited level (~19 cm$^{-1}$), which we calculate using absorption spectra, is close to the averaged energy position of FIR and Raman excitations observed earlier at T>17K, $k$=0, correspondingly, 18 and 22 cm$^{-1}$ [11,12]. These lines were interpreted as Raman and FIR Davydov components of the lowest electron excitation, which are formed in absorption spectra due to the transitions from the ground state to the first excited level of the ground multiplet $^6H_{15/2}$, which is split by the crystal field. Note, that formation of the studied absorption spectra includes the processes in the whole (for $k\neq 0$) Brillouin zone, so the energy intervals, determined from our experiments, may be different from the energy position of the line in FIR and Raman spectra (namely, for $k=0$). But all of them are comparable, and we can conclude, that one-ion consideration is quite correct in our case. More detailed discussion of the nature of absorption spectra fine structure is presented below.

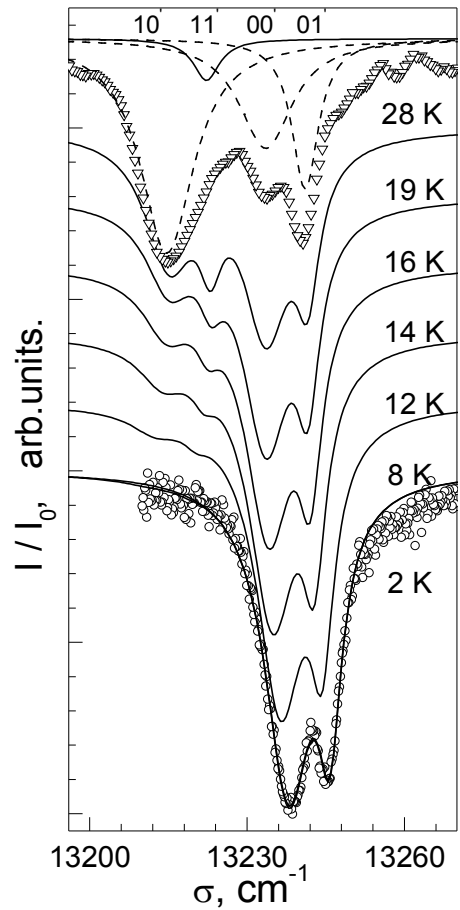

Figure 4. Evolution of the spectra based on the computing: transition $^6H_{15/2} - {}^6F_{3/2}$. Half-width of the initial doublet and additional two lines are linearly changed from its value at 2K. Summary intensity of the spectra is $I_1+I_2$ for all temperatures, here $I_1$ and $I_2$ are intensities of the doublet components at 2K.

Rhombic unit cell contains four molecules {$Dy(MoO_4)_2$}, they transform from one to another by inversion and two - fold axis. So absorption spectrum of the rhombic phase is formed by four rare-earth ions and 2 Davydov's components have to be observed for every excitation. Local symmetry of dysprosium ions is $C_2$ [7], and every Davydov component (corresponding wave functions for electron states) belongs to irreducible representation of $D_{2h}$ point group (isomorphic to $D_{2h}^{14}$ factor-group), odd (⁺) or even (⁻) in respect to inversion. Thus all excited electron states for ions may be (⁺) or (⁻) type, excluding ground state of (⁺) type. Transitions between "odd" and "even" states are dipole active and are presented in the spectrum.

We note here, that dipole light absorption ($k=0$), caused by transitions inside f-configuration of $Dy^{3+}$ ions, is strongly forbidden in the free ions. Since the local point group for $Dy^{3+}$ ions does not include inversion these transitions are not forbidden due to the broken ions symmetry in the rhombic phase unit cell as well as in the low - temperature phase. Further we examine the temperature evolution of the spectra.

At 2K when the lowest excited levels are not thermally occupied, we have observed only two lines, originated from the transition from the ground state to the excited doublet $^6F_{3/2}$. We observe the depolarised Davydov components of the doublet in **E**||**a** and **E**||**c**, and they are not essentially different by energy in two different polarisation. The corresponding splitting is no more then $2cm^{-1}$. In addition to our scheme, a very weak line was observed from the high-energy side of the doublet on the energy distance ~ $22cm^{-1}$.

When the lowest excited electron levels became thermally occupied, a new component of the spectra arises from the low-energy side. This band is formed by transitions from the first excited level (namely, between the corresponding Davydov components of the different parity) and have different energies and intensities. Fine structure spectra, originated from the transitions (0-0', 0-1', 1-0', 1-1'), are determined by different parameters – lattice parameters and lattice symmetry, static and dynamical "odd" contributions to forbidden f-f transitions [13]. These parameters are unknown from independent data, by this reason we cannot calculate correctly the spectra parameters at high temperatures. So we have only fixed their changing at CJTE. The conclusions from the computing, Figure 4, are following.

1. The behaviour of the lines intensity was not described by the factor Bose only and the assumption about the change of symmetry is essential in the region of CJTE.
2. To describe the formation of the absorption from the first excited level at the finite temperature we need for assumption, that the electron-phonon interaction leads to widening spectra, when the electron levels as well as the optical phonon branches are thermally occupied.
3. Two different processes were also observed which were caused by CJTE. At 17K the shift of the lines begins and is finished at ~ 8K. In the region of 12K the kink in the polarisation of the lines was observed, at T<12K the doublet is non-polarised for **E**||**a** and **E**||**c**.

We have used for comparison the data of the temperature behaviour of the rare–earth ions spectra parameters in the "stable" lattice from the literature. It was experimentally shown for $Pr^+$ compounds [13], as the temperature was changed, position of the lines maximum shifted monotonously (practically linearly) in the wide temperature region. The same changes in the line intensities were observed, in contrast to our case. So our conclusion is that the crystal symmetry is lowering due to the phase transformation, and this fact leads to the peculiar change of intensity of some lines, Figure 3.

So, the symmetry of the low-temperature phase is monoclinic, and the active irreducible representations are connected with the zone Billouin boundary point, as it is supposed in [5]. As a result of the symmetry lowering, the expected number of the dipole active Davydov components for the lowest excitations will be twice more than in the initial rhombic phase, because the number of crystallographic non equivalent rare - earth centres in the unit cell is twice more (8 in the low-temperature phase instead of 4). That is why four lines may appear in the low-symmetry phase at least, that form the low - energy side of the spectrum. These lines are weak and are not well resolved at the temperatures 2-12K due to the small differences in Davydov components energies of the first excited level: from [11,12,] and our results, this difference is no more then 2-3cm$^{-1}$ at 2K. It is possible reason, why we didn't observe the resolved doublet structure of the satellite in the studied temperature region. In the next section we return to the peculiarities which were experimentally observed earlier and will discuss the changes of macroscopic features at the phase transitions. Then we shall compare them with our results, and carry out the symmetry analysis.

### 3.2. SYMMETRY CONSIDERATION.

Anomalies of birefringence at CJTE [3,4] in $KDy(MoO_4)_2$ may be caused by the shifts of the energy level positions of the rare-earth ions due to the changes of the local crystal field parameters. During "improper" phase transitions, however, the elastic properties of the crystal, as well as the lattice parameters, change too. Both effects may result in the same picture, observed in the birefringence [3,4]. Further we would discuss the elastic modules anomalies, observed near transition temperatures [7], because we have no data on the lattice parameter behaviour in $KDy(MoO_4)_2$ at CJTE. For convenience we present the fragment of the structure in the Figure 5 and corresponding position of the symmetry elements for $D_{2h}^{14}$ space group.

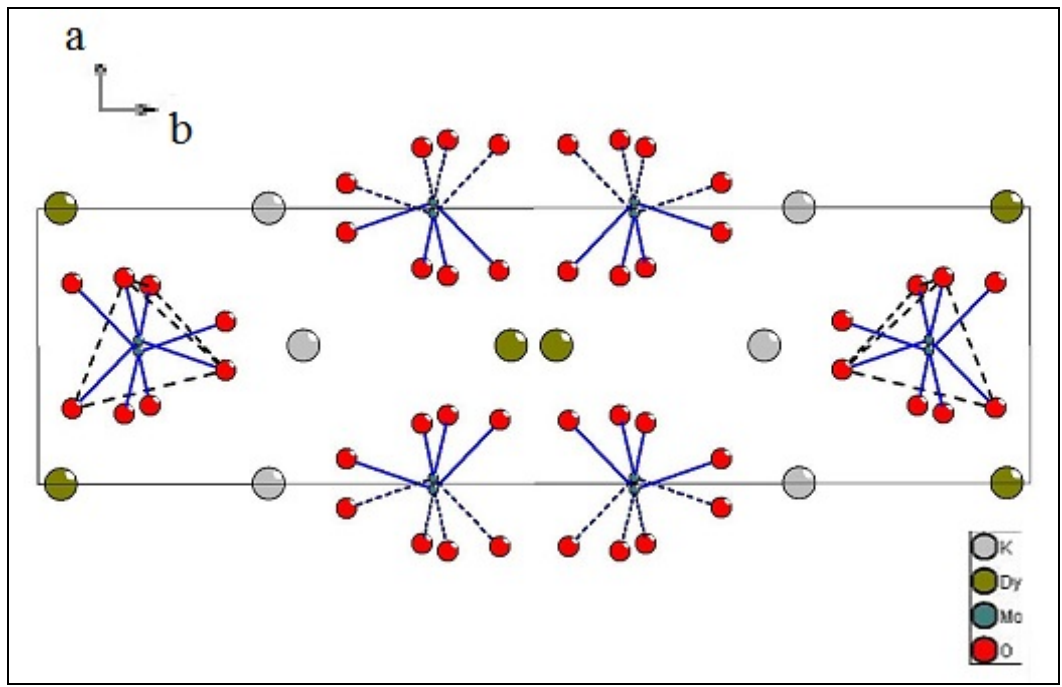

Figure 5. Here $\mathbf{x}\|c$, $\mathbf{y}\|a$, $\mathbf{z}\|b$ are the rhombic crystal axes, $a$ = 5.084A, $b$ =18.18A, $c$ = 7.97A[14].

Elastic module behaviour in the temperature region of CJTE was studied in [6]. Their temperature dependences are shown in Figure 6 (a, b) in the notations corresponding to the elastic energy of a rhombic crystal [5]. We would like to accentuate that we separated curves between two groups to point out that ultrasound anomalies take place at two distinguished "critical" temperatures.

It is seen, that the anomalous temperature dependence of the average interval (0-1') (Figure 4) and the relative intensities (Figure 2) correlate in positions and types with the anomalies in Figure 6, b at $T_1 \sim 17K$ and Figure 6, a at $T_2 \sim 12K$. In [5, 6] it was supposed that at T<12K the spontaneous deformation of $\mathbf{u}_{zy}$ –type appears in the crystal and leads to the monoclinic distortion of the unit cell. Two-fold axis conserves in the layer *ab*, where the rare earth ions are placed, and local $C_2IIy$ in Dy – ion position disappeared [7]. This deformation leads to the depolarization of the spectra at 12 K (Figure 4) due to the small changing of the crystal field parameters and the lowering of crystal field symmetry in the position of rare-earth ions. On the other hand, the temperature region of the energy shift between lines of the spectra correlates with the temperature region of the $c_{11}$ longitudinal modulus changing, which corresponds to $\mathbf{u}_{xx}$ –type deformation of compression (dilatation) (Figure 6, b). That is, spectra fine structure peculiarities are the result of elastic distortions of the different types, appearing at the phase transition in the crystal. Note, that the both types of experiments were performed on different samples and independently, so we believe that our comparison is reliable.

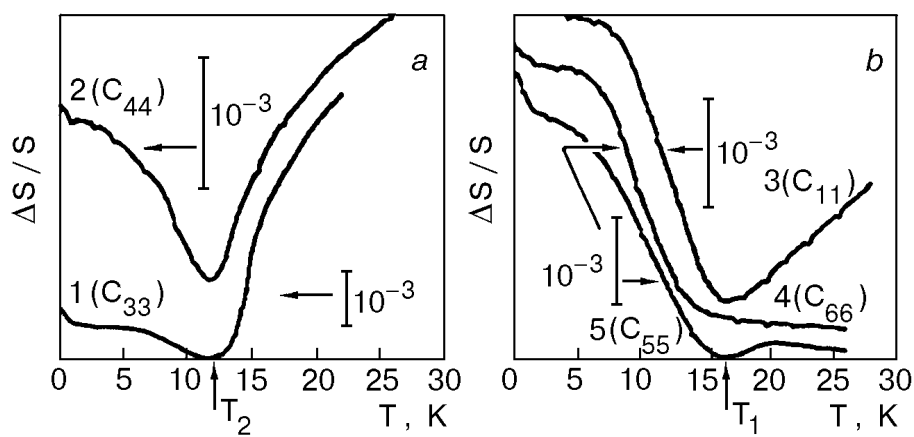

Figure. 6(a, b). Temperature dependences of the longitudinal and transverse sound modes which propagate along the main directions of the rhombic crystal (data were obtained in[9]): 1 - *q* || *b* || *z*; 2 - *q* || *b* || *z*, *u* || *a* ||*y*; 3 - *q* || *c* ||*x*; 4 - *u* || *c* || *x*; *q* || *a* || *y*; 5 - *q* || *b* || *z*; *u*||*c* ||*x* . The scales of the $C_{66}$ and $C_{55}$ modules are the same.

Really, wave functions of the ground multiplet $^6H_{15/2}$ in the free ions belong to the double-degenerate representation $E'_{1/2}$ of the double group [15]. In the rhombic crystal field the ground multiplet is split on 8 Kramers doublets. Wave functions for every Kramers doublet are linear combination of the spherical harmonics with half – integer values of the full moment J = 15/2, 13/2, 11/2 and so on, and are characterized by the same type of symmetry. Direct production of $E'_{1/2} *E'_{1/2}$ representations contains both A and B representations of the local point group $C_2$. So every distortions belonging to $A_g$, $A_u$, $B_{1g}$, $B_{1u}$ and other reperesentations of the rhombic factor-group $D_{2h}$

(see Table), may affect the position of the electron states as well as the polarisation feature of the spectra fine structure. Since the ultrasound anomalies are quite diffuse, we nevertheless use some conclusion from the symmetry approach for discussion.

Different features of ultrasound anomalies at 12K and 17K were discussed in [8,10]. Based on the modules behavior at $T_1$ and $T_2$ it was supposed that phase transformations in $KDy(MoO_4)_2$ are caused by different active zone Brilllouin points of the initial $D_{2h}^{14}$ space group. Double-degenerate irreducible representations, connected with $k_{23} = 1/2(b_2+b_3)$, may induce "improper" ferroelastic phase transition of displacement type to commensurate monoclinic phase [5,6]. As a result of the phase transition, the week softening of the modulus $c_{44}$ takes place, which as supposed is the typical for second order phase transition. Arising of $c_{11}$ modulus, beginning at 17K is similar to anomaly shown on Figure 2, which permits us to suppose that the other point is active also in the "pre-transition" region, and only $k_{19} = 0$ may be considered [8]. The only possible case is described further.

We excluded the activity of $B_u$ and $B_g$ - types representations at CJTE because phase transition in $KDy(MoO_4)_2$ is not "pure" ferroelectric or ferroelastic one, and choose the order parameter $\xi$, which transforms as $A_u$ representation ($\tau_2$ by [9]) of $D_{2h}$ point group as the most realistic. $A_u$ representation leads to the second order phase transition and induces rhombic phase $D_2$ (Z=4, as in the initial rhombic phase). In this case shear modules will not be softening near phase transition temperature, in contrast to the case of "improper" ferroelastic phase transitions to monoclinic phases. Dielectric constant anomalies at the phase transition $D_{2h}$ - $D_2$ should be very weak because of the fact that the non – centro-symmetric phase $D_2$ is non-ferroelectric one. At the rhombic - rhombic phase transition only jumps of the longitudinal modules are expected. As it seemed (see above), arising of $c_{11}$ contradicts to the conclusions of the previous discussion. Let us consider, however, the possible physical origin of the order parameter $\xi$, taking into account, that the arising of $c_{11}$ and energy shifts in the spectra are correlated. Further we will discuss, what types of the low-energy phonons are possible for rhombic $KDy(MoO_4)_2$.

It is known that the expansion of mechanical representation on the irreducible ones gives us:

$$\Gamma_N = 17 A_g + 19 B_{1g} + 19 B_{2g} + 17 B_{3g} + 17 A_u + 19 B_{1u} + 19 B_{2u} + 17 B_{3u}, \qquad (2)$$

here $A_g$, $B_{1g}$ and so on are the irreducible representations of $D_{2h}$ factor-group, and coefficients in (2) show the number of the optical phonon modes, whose symmetry coordinates transform as corresponding representations[14].

As it can be shown, $A_u$ representation includes 4 internal degrees of freedom - 3 translation and 1 rotation ones. They describe shifts of different ions and rotations of $(MoO_4)^{2-}$ tetrahedrons. We think that these phonon modes have low energies in $k_{19} = 0$ which are close to the low-energy side of the optical phonon excitations.

For further discussion note, that $KDy(MoO_4)_2$ crystal may be presented as a layered structure with the layers of $\{(Dy(MoO_4)^2\}^-$ (in the plane **ac**). $KDy(MoO_4)_2$ includes two similar layers in the rhombic unit cell which are mutually shifted along plane. It was shown [11], that the low-energy side of the phonon excitations may be described as the result of the "frozen" mutual shifts of the layers $\{(Dy(MoO_4)^2\}^-$ in the two-layered crystal and these excitations are active in FIR region, because inversion is conserved in the layers.

Lattice modes which correspond to $A_u$ modes are active neither in Raman nor IR spectrum. Wave functions transformed by $A_u$ matrix may be constructed as direct productions of $B_g$ and $B_u$ - types rotation or translation modes:

$$A_u = B_{3g}*B_{3u} \t{6} \quad (3)$$
$$= B_{2g}*B_{2u},$$
$$= B_{1g}*B_{1u}.$$

Introduce the shifts $(x_I - x_{II})$ of neighboring layers along y II **a** with the numbers I and II. These displacements belong to $B_{1u}$ representation, because two-fold axis $C_2IIx$ is conserved, when alternative shifts of the layers take place. The symmetry coordinates constructed as a direct production of $(x_I - x_{II})$ and $u_{yz}$

$$(x_I - x_{II})* u_{yz}, \quad (4)$$

belong to $A_u$ irreducible representation. We may include also in the consideration the parameter $R_y = (R_1 + R_2) - (R_3 + R_4)$, where 1,2 is used for JT center in the layer I and 3,4 - in the neighboring II one – which transforms as $B_{2g}$ or as axial vector $R_y$. It may be considered as the rotation (conditionally) around $C_2IIy$ of the oxygen polyhedron surrounding of JT ions and formed by the $(MoO_4)^{2-}$ tetrahedrons so the production belongs to $A_u$ too. However we don't construct the corresponding symmetry coordinate, now we only supposed, that the shifts of the neighboring layers and rotations of tetrahedral, surrounding of JT ions in the different neighboring layers, correspond with $A_u$ type vibrations. Otherwise, we suppose that shifts of the layers and rotations of the tetrahedral are coupled. This coupling is a result of the specific "skeleton" type structure $KDy(MoO_4)_2$ [16].

Now let us discuss the physical nature of $A_u$ modes. Its energy may be close to zero or have the gap as an optical mode. The lattice internal modes, translation and rotation ones, energies in

KDy(MoO$_4$)$_2$ lay in the energy region 100 - 200cm$^{-1}$ [16] for KDy(MoO$_4$)$_2$, but the "mixed" modes, which have to be active in CJTE transition as supposed, may be some lower by energy than initial internal "pure" modes. On the other side, when the softening of the low-energy zone boundary phonon branch at T >~T $_{CJTE}$ is beginning, all acoustic modules near $k$ =~ 0 due to the phonon interaction are softening too.

We follow now, that the splitting of the acoustic spectra in two-layered crystal leads to appearance of the optical dipole active branches in KDy(MoO$_4$)$_2$ with energy positions 20 and 28 cm$^{-1}$, correspondingly, and its energies in $k_{19}$ =0 are determined by the "shear-type" $c_{55}$ and $c_{44}$ modules. So, when the layers vibrate, the A$_u$ and B$_u$ -type modes would be excited at the energies close to the low-energy side of the optical phonon spectra (it follows from the form of symmetry coordinate (4)).

### 3.3 NATURE OF CJTE in KDy(MoO$_4$)$_2$.

Vibrations of (MoO$_4$)$^{2-}$ complexes, accompanied with the layers shifts, modulate the crystal field parameters in JT ions positions. If A$_u$ mode is active at CJTE, the corresponding shifts and corresponding rotations are "frozen" and the energy levels positions are changed. We supposed, that A$_u$-type "coupled" modes exist in the two-layered crystal really. We keep in mind a specific effect of the external magnetic field on Raman active excitation at 23 cm$^{-1}$ and 32cm$^{-1}$ (at T>T$_{cr}$ only one low - energy excitation was observed – about 20cm$^{-1}$) [10]. In addition, A$_g$ type excitations appeared in Raman spectra at T<<T$_{cr}$, because A$_u$ modes became Raman active due to the loss of the symmetry and conserved C$_2$IIc at the low temperatures. At T<<T$_{cr}$ every line is split on two components under external magnetic field (up to 6T), and one of the component does not shift under magnetic field. Non-shifted components exist because the modes are not "pure" rotation or translation, or "pure" electron or phonon ones. So we think, that A$_u$ mode is the "mixed" phonon-electron mode, which couples electron wave function for the first electron level with the displacement of the layers.

Really, if we include in the consideration the interaction of the electron states with the low-energy optical phonon branch only, i.e. with the shifts of the layers without the deformation of JT center, we cannot explain the large change of the orientation of g-factor axes [6,7]. In the case of A$_u$-type mode activity at CJTE, when rotations of tetrahedron are "frozen", the orientation of magnetic spin moments of the Dy$^{3+}$ ions are effectively changed and the CJTE critical temperature is operated by external magnetic field [4]. Note, that the "freezing – up" of A$_u$ – type mode symmetry coordinates may lead to the changing of the crystal size along direction x (because of the shifts +x and -x of the all neighbor layers) as well as the appearing of the shift-like deformation $u_{yz}$. Really, Figure 6, b shows, that at T~17K anomalous arising of $c_{11}$ and small "softening" of the $c_{55}$ elastic modules were observed. We can suggest also, that the arising of $c_{11}$ is occurred simultaneously with the changes of the crystal sizes. But only $A_u$ mode activity does not describe the anomalous behavior of the other modules at 12K. We think, that the "frozen" layer shifts, accompanied with the rotation-like vibrations of the oxygen tetrahedron, lead to spontaneous deformation at phase transition to monoclinic phase at 12K, but in the way, which is not so simple.

To discuss the possible nature of the CJTE and intermediate phase between 17K and 12K let us introduce a quasi-degeneration mode with the components ξ and p, which transforms as $A_u$ and $B_{2g}$, representations, correspondingly, and consider the two-component order parameter (ξ, p). Thermodynamic potential Ψ(ξ, p) as a function of two-dimensional order parameter is constructed in the frame of Landau theory taking into account the irreducible representation

```
  E    2_x(τ_x,τ_y,0) 2_y(τ_x,τ_y,0)  2_z    i(τ_x,0,τ_z)  σ_x(0,τ_y,τ_z) σ_y(0,τ_y,τ_z) σ_z(τ_x,0,τ_z)
_________________________________________________________________________________________________

  1 0    1 0          1 0             1 0    -1 0          -1 0           -1 0            -1 0         ξ
  0 1    0 -1         0 1             0 -1   0 1           0 1            0 -1            0 -1         p
_________________________________________________________________________________________________
```

As it follows from the ordinary symmetry consideration, three low-symmetry commensurate phases of different symmetry are possible, induced by thermodynamic potential: I – $D_2$ (parameter ξ), II – $C_{2h}$ (parameter p), and III – $C_2$ or $C_s$ (parameters ξ and p) for all cases (3). In this case, the thermodynamic function Ψ(ξ, p) includes the gradient invariants of Lifshits type [17] (in contrast to the case of one-dimensional representations ξ and p). Let us construct the invariant,

$$p * d\xi/dx - \xi * dp/dx , \qquad (7)$$

here ξ transforms as $A_u$, x - as $B_{2u}$, p- as $B_{2g}$. Expression (6) is invariant form for $D_{2h}$ point group and must be added to the thermodynamic potential Ψ( ξ, p ). Note, that the gradient terms

$$p * d\xi/dy - \xi * dp/dy , \quad p * d\xi/dz - \xi * dp/dz \qquad (8)$$

are possible for $D_{2h}$ group by formal reason too [17]. Because of the existence of the gradient invariants, the two-dimensional order parameter (ξ, p) induces the incommensurate phases (IP), which "precedes" the commensurate (monoclinic, for example) ones, determined above.

If the point $k_{19} = 0$ is active only, the unit cell is not doubled at T<12K. The monoclinic axis $C_2 \text{II} \textbf{c}$ in this case (7) is the same as in the previous case, but incommensurate modulation will appear along *c* axis with (*k*, 0, 0), where parameter *k*, *k* II x II *c*. The temperature region of IP existence is determined by the coefficients of the thermodynamic potential, which includes the gradient terms. So IP may exist in the phase down to the lowest temperatures.

Note, that it is the simplest approach. There are more complicated cases, which are not distinguished from the first one, by ultrasound anomalies, at least. Let us suppose, that the order parameters, connected with $k_{23} = 1/2(\textbf{b}_2+\textbf{b}_3)$, are the "main" parameters. The nonlinear couple of the active parameter and parameter ξ "switch" new phase, namely, IP one, and this phase exists in the temperature region between 17K and 12K (from the birefringence measurements, this temperatures are a little different). In other words, we conclude, that the incommensurate phase is induced by JT-type interaction in the "pre-transition" temperature region. Without JT interaction only commensurate phase may exist, because $k_{23}$ activity leads to the commensurate monoclinic phase with $C_2 \text{II} \textbf{c}$ and to doubling of the unit cell volume.

## 4. CONCLUSIONS

Correlation in the changing of the absorption spectra and ultrasound temperature anomalies makes it possible to discuss the physical nature and symmetry aspects of the Jahn-Teller type ordering in $KDy(MoO_4)_2$. The main conclusions are the following.

We have observed two temperature regions of CJTE in absorption spectra, which are determined by two-stage ordering in $KDy(MoO_4)_2$. $A_u$ – type electron-phonon mode, constructed as the combination of the alternative shifts of the neighboring layers and the distortions of the polyhedrons around JT ions, is active at CJTE and may lead to incommensurability in the intermediate temperature region 17K - 12K near *k* ~ 0. We may conclude, moreover, that the lowering of the ground state of the dysprosium subsystem leads to the anomalous elastic energy changing accompanied with the arising of $c_{11}$ and $c_{55}$ modules at 17K. At 12K the low - temperature monoclinic phase is stabilized and the unit cell doubling takes place, accompanied with the lowering of the position symmetry of rare-earth ions.

# REFERENCES

[1] A.I. Zvyagin, T.S. Stetsenko, V.G.Yurko and R.A. Vayshnoras, *Pisma v ZhETF.,* **17**, 90(1973).
[2] M.J.M. Leask, O.C. Tropper and M.L. Wells, *J. Phys. C: Solid State Phys.* **14**, 3481(1981).
[3] Yu.N. Kharchenko, *Low Temp.Phys.,* **22**, 382(1996).
[4] N.F. Kharchenko and Yu.N. Kharchenko, R. Szymczak and M. Baran, *Low Temp.Phys.,* **24**, 989 (1998).
[5] I.M. Vitebski, S.V. Zherlitsyn, A.A. Stepanov and V.D. Fil'. L*ow Temp.Phys*, **16**(8), 1064(1990).
[6] I.M. Vitebski, S.V. Zherlitsyn, A.I. Zvyagin, A.A. Stepanov, V.D. Fil', *Low Temp.Phys.,* **12**, 1108 (1986).
[7] V.A. Bagulya, A.I. Zvyagin, M.I. Kobets, A.A. Stepanov, A.S. Zaika, *Low Temp. Phys.,* **14**, 493(1988).
[8] N.M. Nesterenko, *Physics of the Solid State,* **42**(1)**,** 184 (2000).
[9] O.V. Kovalyev. *Irreducible representations of the space group*, in Russian ( Izd-vo: Akad.Nauk Ukrainy, Kiev ,1961).
[10] Ya.I. Zagvozdina, N.M. Nesterenko and Yu.N. Kharchenko, *Ferroelctrics*, **239**, 197(2000).
[11] D. Mihailovic, J. F. Ryan, M.C.K. Wiltshire, *J. Phys. C: Solid State Phys.* **20**, 3047(1987); ibid, 3063(1987).
[12] V.I. Kut'ko, Yu.N. Kharchenko, N.M. Nesterenko and A.A. Gurskas, *Low Temp. Phys.,* **22**, 603(1996).
[13] I.S. Édelman, A.V. Malakhovskii, A.M. Potseluyko, T.V. Zarubina, A.V. Zamkov, *Physics of the Solid State*, **43**, 1037-1042(2001).
[14] V.A. Vinokurov and P.V. Klevtsov. *Kristallografiya*, **17**, 127(1972).
[15] V.V. Eremenko. Introduction to magnetic spectroscopy (in Russian). "Naukova dumka", Kiev, 1975. 472 p.
[16] J. Hanuza and L. Macalik, *Spectrochimica Acta,* **38A**, 61(1982).
[17] P. Kuzel, P. Moch, A. Gomez-Cuevas, V. Dvorak, *Phys.RevB* **49,** 6553(1994).